\documentclass[prl, twocolumn]{revtex4-1} 

\usepackage{hyperref}
\usepackage{amsmath}
\usepackage{balance}
\usepackage{graphicx}
\usepackage{color}
\usepackage{setspace}

\begin{document}

\title{The interaction of Kerr nonlinearity with even-orders of dispersion: an infinite hierarchy of solitons}


\author{Antoine F. J. Runge$^{1,*}$}
\author{Y. Long Qiang$^{1}$}
\author{Tristram J. Alexander$^{1}$}
\author{Darren D. Hudson$^{2}$}
\author{Andrea Blanco-Redondo$^{3}$}
\author{C. Martijn de Sterke$^{1,4}$}

\affiliation{
$^{1}$Institute of Photonics and Optical Science (IPOS), School of Physics, The University of Sydney, NSW 2006, Australia\\
$^{2}$CACI-Photonics Solutions, 15 Vreeland Road, Florham Park, NJ 07932, USA\\
$^{3}$Nokia Bell Labs, 791 Holmdel Road, Holmdel, NJ 07733, USA\\
$^{4}$The University of Sydney Nano Institute (Sydney Nano), The University of Sydney, NSW 2006, Australia\\
$^{*}$Corresponding author: antoine.runge@sydney.edu.au
}

\begin{abstract}
Temporal solitons are optical pulses that arise from the balance of negative group-velocity dispersion and self-phase modulation. For decades only quadratic dispersion was considered, with higher order dispersion thought of as a nuisance.
Following the recent reporting of pure-quartic solitons, we here provide experimental and numerical evidence for an infinite hierarchy of solitons that balance self-phase modulation and arbitrary negative pure, even-order dispersion. Specifically, we experimentally demonstrate the existence of solitons with pure-sextic ($\beta_6$), -octic ($\beta_8$) and -decic ($\beta_{10}$) dispersion, limited only by the performance of our components, and show numerical evidence for the existence of solitons involving pure $16^{\rm th}$ order dispersion. Phase-resolved temporal and spectral characterization reveals that these pulses, exhibit increasing spectral flatness with dispersion order. The measured energy-width scaling laws suggest dramatic advantages for ultrashort pulses. These results broaden the fundamental understanding of solitons and present new avenues to engineer ultrafast pulses in nonlinear optics and its applications.
\end{abstract}

\maketitle

\section*{Introduction}
Solitons are among the most striking phenomena in nonlinear physics and have been observed in a wide range of systems \cite{Polturak_1981, Denschlag_2000}. In optics, these transform-limited, shape-maintaining pulses have been crucial in the development of numerous applications ranging from telecommunications \cite{Mollenauer_1991, Haus_1996}, to frequency comb generation \cite{Herr_2014, Yi_2015} and mode-locked lasers \cite{Mollenauer_1984, Zhou_1994}. Traditionally, the formation of these wavepackets relies on the balance between self-phase modulation (SPM) and negative quadratic dispersion ($\beta_2 < 0$), while higher dispersion orders were seen as a nuisance, leading to the emission of dispersive waves in fibers \cite{Elgin_1992, Kodama_1994} and laser cavities \cite{Dennis_1994, Santagiustina_1997}, or acting to limit the achievable pulse duration \cite{Hook_1993, Chan_1994, Aceves_1994, Christov_1994}.

When a pulse propagates through a nonlinear medium, SPM causes the generation of new low frequencies on the leading edge of the pulse, and new high frequencies on its trailing edge. To understand solitons, recall that in the presence of pure negative dispersion $\beta_k$ of order $k$ at frequency $\omega_0$, the inverse group velocity $v_g$ for frequencies close to $\omega_0$ can be written as 
\begin{equation}
    {1\over v_g}={1\over v_{g0}}-{1\over{(k-1)!}} |\beta_k|\, (\omega-\omega_0)^{k-1}.
\label{eq:vg}
\end{equation}
Here $v_{g0}$ is the group velocity at $\omega_0$, and $k=2$ for quadratic dispersion, etc. For negative quadratic and in fact for all high, even-order types of dispersion, \eqref{eq:vg} shows that the group velocity monotonically increases with frequency. Consequently, both the SPM-generated low frequencies on the leading edge and the SPM-generated high frequencies on the trailing edge move towards the pulse center, leading to the formation of a soliton. This argument suggests that temporal solitons should exist in the presence of any negative even-order dispersion. Indeed, recent studies showed that optical solitons could arise rom the balance between SPM and negative quartic dispersion ($\beta_4 < 0$) \cite{Blanco_Redondo_2016,Tam_2019, Runge_2020}. However, the existence of solitons for higher, even dispersion order ($k > 4$) and their properties are yet to be reported.

Here we report experimental evidence for an infinite hierarchy of solitons, to which we refer as pure high, even-order dispersion (PHEOD) solitons, arising from the balance between SPM and any negative even-order of dispersion, of which conventional solitons and pure-quartic solitons are its two lowest-order members. We experimentally demonstrate three new members of this hierarchy, namely pure-sextic, -octic and -decic solitons (arising, respectively, from the balance of SPM and negative dipsersion of order $k=6$, $8$, and $10$). The experimental results agree well with numerical results found by solving the nonlinear Schr\"{o}dinger equation (NLSE), modified to higher orders of dispersion. In fact, in this way we provide evidence PHEOD solitons of order $16$. 

While previous theoretical works have studied the combined effects of high-order dispersion of order up to $k=8$ and nonlinear effects (see, e.g., \cite{Kedz2015,PhysRevE.93.012206,Sun:16,JIA201790}), they all consider equations with a large number of terms, many of which representing combined nonlinear and dispersive effects. Our aim differs in that we consider the Kerr nonlinear effect and pure, even, high-order dispersion (see \eqref{NLSE} below).

The PHEOD solitons are generated in a passively mode-locked laser with tunable net-cavity dispersion \cite{Runge_2020}. The solitons have spectral sidebands, associated with resonant dispersive waves typical of fiber lasers, the spacing of which is directly, and quantitatively, related to the dispersion. By measuring the PHEOD solitons' energy we find that they are related to the pulse duration, $\tau$, as $E \propto \tau^{-(k-1)}$. The strongly increasing pulse energy with decreasing pulse length demonstrates the potential of high-order dispersion for unlocking innovations in nonlinear optics. Our results establish a new degree of freedom for the generation and study of ultrashort optical pulses with potential applications in lasers \cite{Runge_2020} and frequency comb generation \cite{Bao_2017, TaheriMatsko_2019}.

\section{Numerical results}

We consider the propagation of optical pulses in a medium with Kerr nonlinearity and $k^{\rm th}$ order dispersion, where $k$ is an even integer. This evolution can be described by the modified NLSE 
\begin{equation}
   i\frac{\partial \psi}{\partial z}=-(i)^k\frac{|\beta_k|}{k!}\frac{\partial^k \psi}{\partial T^k} - \gamma |\psi|^2\psi,
    \label{NLSE}
\end{equation}
where $\psi(z,T)$ is the pulse envelope, $z$ is the propagation coordinate, $T$ is the local time, $\beta_k$ is the dispersion coefficient, which is taken to be negative, and $\gamma$ is the nonlinear parameter. For $k > 2$, \eqref{NLSE} is non-integrable and has no known analytic pulse-like solutions. However, we can look for stationary solutions of \eqref{NLSE} of the form $\psi(z,T) = A(T;\mu)e^{i\mu z}$,
which satisfy 
\begin{equation}
\mu A-(i)^k\frac{|\beta_k|}{k!}\frac{\partial^k A}{\partial T^k}- \gamma A^3=0,
\label{eq:A}
\end{equation}
so the shape is preserved during propagation, and $A$ can be taken to be real.  We solve~\eqref{eq:A} using the Newton-conjugate-gradient method \cite{Tam_2019, Yang_2009}.

\begin{figure}[hbt]
\centering
\includegraphics[width=90mm,clip = true]{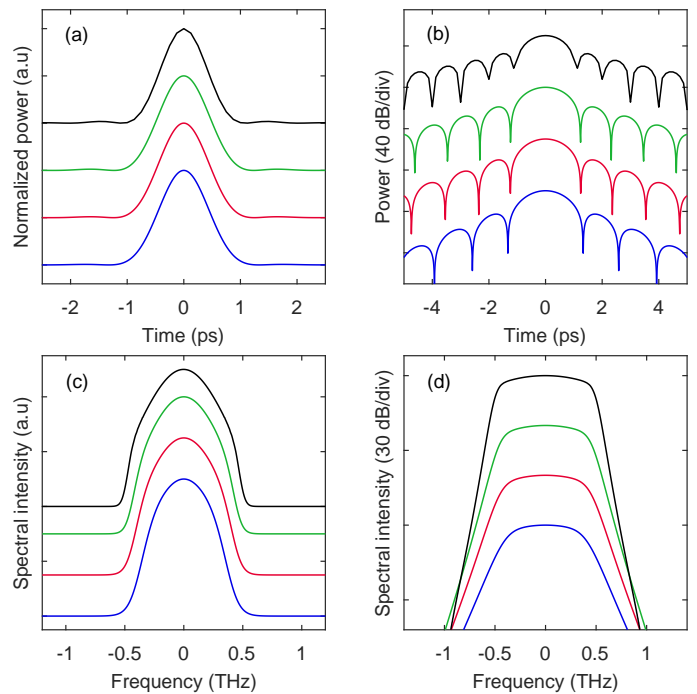}
\vskip 0mm
\caption{Numerically calculated temporal stationary solutions for $k = 6~\rm{(blue)},~8~\rm{(red)},~10~\rm{(green)}~\rm{and}~16~\rm{(black)}$ with a same pulse width (FWHM) of $\tau = 1~\rm{ps}$. Temporal profiles in linear (a) and logarithmic (b) scales. Corresponding linear (c) and logarithmic (d) spectra. The different solutions have been shifted vertically for clarity.}
\label{num_sol_fig}
\vskip-1mm
\end{figure}

\begin{figure*}[hbt]
\centering
\includegraphics[clip = true]{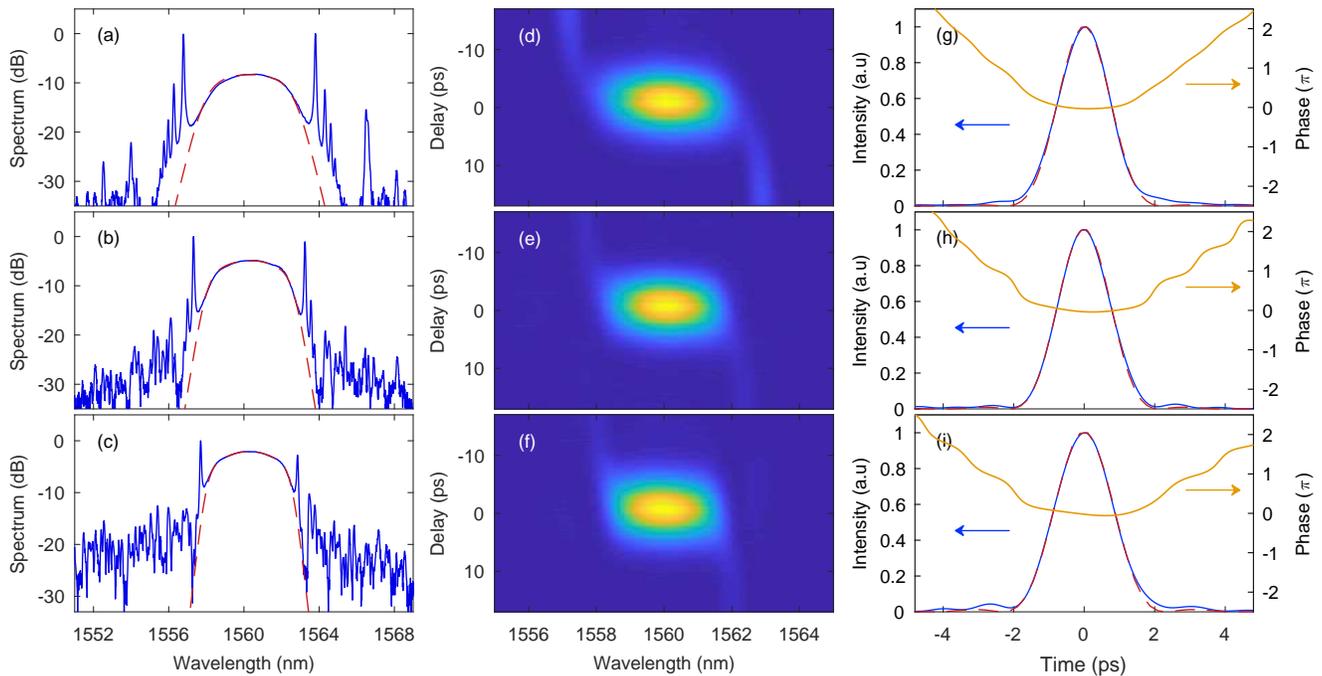}
\vskip 0mm
\caption{Spectral and temporal measurements of pure high-order dispersion solitons sextic (top row), octic (middle row) and decic (bottom row) dispersion. The applied dispersion is $\beta_6 = -500~{\rm ps^6/km}$, $\beta_8 = -15\times 10^3~{\rm ps^8/km}$ and $\beta_{10} = -1\times 10^6~{\rm ps^{10}/km}$, respectively. (a)-(c) Measured (blue), and calculated (red-dashed) spectra. (d)-(f) Measured spectrograms. (g)-(i) Retrieved temporal intensity (blue), phase (orange) and corresponding calculated temporal shapes (red-dashed).}
\label{laser_out}
\vskip-1mm
\end{figure*}

Solving \eqref{eq:A} provides a single solution for each dispersion order, but using a scaling argument we can obtain an entire family of solutions. Since \eqref{NLSE} is invariant under the transformation 
\begin{align}
  A &\rightarrow \alpha A & \tau &\rightarrow \alpha^{-2/k}\tau & \mu &\rightarrow \alpha^2\mu,
  \label{scaling_eq}
\end{align}
there exists a continuous family of PHEOD solitons for each dispersion order $k$, each with the same pulse shape, but with different amplitudes and widths, that can be parameterized by their values of $\mu$ \cite{Tam_2019}. From~(\ref{scaling_eq}) it can be shown straightforwardly that the energy $E_k$ of a soliton with pure even dispersion order $k$ scales as $E_k\propto\tau^{-(k-1)}$.

The numerically calculated temporal and spectral intensity profiles of the resulting  stationary solutions for $k = 6,~8,~10~\rm{and}~16$, with the same temporal pulse duration at full width half maximum (FWHM) $\tau = 1~\rm{ps}$, are shown in Fig.~\ref{num_sol_fig}. We note that similar to PQSs, the pulse temporal shapes of the PHEOD solitons exhibit oscillations in the  tails \cite{Tam_2019, Akhmediev_1994}, which become more prominent with increasing dispersion order, as seen in Fig.~\ref{num_sol_fig}(b). Simultaneously the central part of the associated spectrum becomes increasingly flat (see Fig.~\ref{num_sol_fig}(c) and particularly in  Fig.~\ref{num_sol_fig}(d)). This increased spectral flatness can qualitatively be understood from \eqref{eq:vg} for the group velocity near frequency $\omega_0$. It shows that for high dispersion orders $k$ the group velocity remains approximately constant around $\omega_0$ before changing rapidly. There is thus a frequency interval around $\omega_0$ for which the dispersion is essentially irrelevant, and in this frequency interval the spectral intensity does not need to vary significantly to balance the dispersion. To understand the effect of the spectral flatness in the time domain, recall that the second derivative of a function corresponds to the second moment of its Fourier transform. Thus the Fourier transform of a flat function, i.e., a function with a small second derivative, must have a small second moment, which can only be achieved by sign changes. For the pulses we are considering this corresponds to oscillations in time \cite{Tam_2019}. 

\section{Experimental setup and results}

The intrinsic dispersion of conventional optical waveguides is dominated by quadratic contribution ($\beta_2$) while the effects of higher-order dispersion are usually weak. In fact a complex structure was required just to achieve dominant negative quartic dispersion \cite{Blanco_Redondo_2016, Lo_2018}. This strongly limits the possibility of observing higher-order dispersion soliton propagation in waveguides. To overcome this limitation, and achieve the dominant negative high-order dispersion required for the generation of these novel solitons, we used a passively mode-locked fiber laser similar to the one reported by Runge {\sl et al.} \cite{Runge_2020}. The laser incorporates an intracavity programmable spectral-shaper, which is used to adjust the net-cavity dispersion \cite{Runge_2020, Schroder_2010, Peng_2016}. The applied phase mask compensates for the intrinsic second, third and fourth order dispersion of the fiber components, and applies a large negative, high even-order dispersion. The applied phase profile can be written as
\begin{equation}
   \phi(\omega) = L\left(\sum_{n=2}^{4}{\frac{\beta_n(\omega-\omega_0)^n}{n!}} + \frac{\beta_k(\omega-\omega_0)^k}{k!}\right),
\label{phase_eq}
\end{equation}
where $L = 18.17~{\rm m}$ is the cavity length, $\beta_n$ is the $n^{th}$ dispersion order for $n=2,~3~\rm{and}~4$ to account for the cavity dispersion. For the results presented in this work $\beta_2 = +21.4~{\rm ps^2/km}$, $\beta_3 = -0.12~{\rm ps^3/km}$, and $\beta_4 = +2.2 \times10^{-3}~{\rm ps^4/km}$. These values are chosen to compensate for the dispersion of the SMF used in our setup and are based on values reported in \cite{Hammani_2011, Ito_2016}. The second term on the right-hand side of \eqref{phase_eq} corresponds to the negative high, even-order dispersion required for the generation of sextic ($k = 6$), octic ($k = 8$) or decic ($k = 10$) PHEOD solitons. To obtain the complete spectral and temporal characterization of the pulses, we used a frequency resolved electrical gating (FREG) setup which allows for the measurement of the pulse spectrogram \cite{Dorrer_2002, Blanco_Redondo_2016}. The temporal intensity and phase of the pulses are then retrieved using a conventional blind deconvolution numerical algorithm \cite{Trebino_1997}. 

\subsection{Spectral and temporal characterization}
\label{sec:spatio}

The results of the spectral, temporal and phase-resolved measurements of the output pulses for the laser operating with pure-sextic, octic and decic dispersion are shown in the first, second and third rows of Fig.~\ref{laser_out}, respectively. Figure~\ref{laser_out}(a)-(c) (left column) show the measured output spectra (blue curves) and the corresponding numerically calculated pulse shapes (red-dashed curves), for the three different dispersion orders. The measured spectral $-3~\rm{dB}$ bandwidths are $\Delta\lambda = 3.9,~3.8~\rm{and}~3.8~\rm{nm}$, for the sextic, octic and decic PHEOD solitons, respectively. For all cases the experimental and predicted spectra agree well. We note that the spectral fluctuations away from the pulses arise from the limited spectral resolution of the pulse-shaper. For rapidly varying functions, it undersamples the phase profile leading to aliasing in the applied phase mask, particularly for the highest dispersion order for which the phase varies rapidly for frequencies away from $\omega_0$. Since the spectral fluctuations thus appear far from the central frequency and at least $10~\rm{dB}$ below the peak (see Fig.~\ref{laser_out}(c)), we are confident that they do not affect the pulse dynamics significantly.

This assertion is confirmed by the corresponding measured spectrograms in Fig.~\ref{laser_out}(d)-(f), which show clear unchirped pulses for all three cases. The vertical streaks at short and long wavelengths correspond to the first sidebands \cite{Kelly_1992}. Finally, the temporal intensity and phase profiles of the sextic, octic and decic PHEOD solitons are shown in Fig.~\ref{laser_out}(g)-(i), respectively. The retrieved FWHM pulse durations of the pure-sextic, -octic and -decic solitons are $\tau = 1.68,~1.69~\rm{and}~1.77~\rm{ps}$, respectively. For all cases, the measured temporal intensities (blue curves) are in good agreement with the corresponding numerical solutions (red-dashed curves) for similar FWHM. The retrieved temporal phase (orange curve) indicates that the emitted pulses are slightly chirped. This is because our experimental setup is a lumped system in which the required dispersion is applied at a single point in the cavity, just before the output coupler \cite{Runge_2020}. Note that the numerically predicted oscillations in the tails of the temporal profiles (see Fig.~\ref{num_sol_fig}(b)) are not observed since these are expected to appear approximately $20~\rm{dB}$ below the pulse's maximum which is below the background in our experiments.

\begin{figure}[hbt]
\centering
\includegraphics[width=90mm, height=4cm]{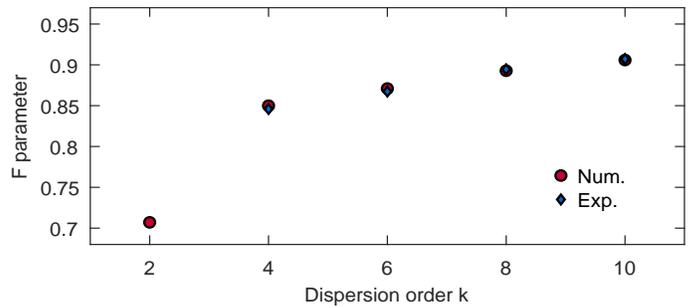}
\vskip 0mm
\caption{Numerical (red circles) and experimental (blue diamonds) values of the flatness $F$ of PHEOD soliton spectrum versus dispersion order.}
\label{bluntness_fig}
\vskip-1mm
\end{figure}

As discussed in Section~3\ref{sec:spatio}, the central part of the PHEOD soliton spectrum becomes increasingly flat with increasing order of dispersion. The flatness or peakedness of a function is often expressed in terms of the kurtosis, but this measure has recently been discredited \cite{Westfall_2014}. Instead we introduce the flatness $F$, which we define to be the fraction of the pulse energy that is within its spectral FWHM. Since it is a fraction, $F$ is intrinsically normalized: it has maximum $F=1$ for a rectangular function and $0<F<1$ for all other functions. The flatness $F$ has the additional advantage that it is an intrinsic property of a function and does not depend on its parameters. For example, for all Gaussian functions $F={\rm erf}(\sqrt{\ln2})\approx0.761$, where ${\rm erf}(x)$ is the error function, and $F=1/\sqrt2\approx0.707$ for all squared hyperbolic secants. The circles in Fig.~\ref{bluntness_fig} give the flatness $F$ of the numerically calculated spectra from Fig.~\ref{num_sol_fig} for different even dispersion orders $k$. The measured values of $F$ (blue diamonds) from the spectra from Fig.~\ref{laser_out} and Ref.~\cite{Runge_2020}, agree to the numerically calculated within $2\%$. This confirms that the flatness of the spectrum increases monotonically with $k$. It has been shown that flatter spectra can lead to enhanced pump-comb conversion and smaller line-to-line power variations in frequency combs \cite{TaheriMatsko_2019}.

\begin{figure}[hbt]
\centering
\includegraphics[width=90mm,clip = true]{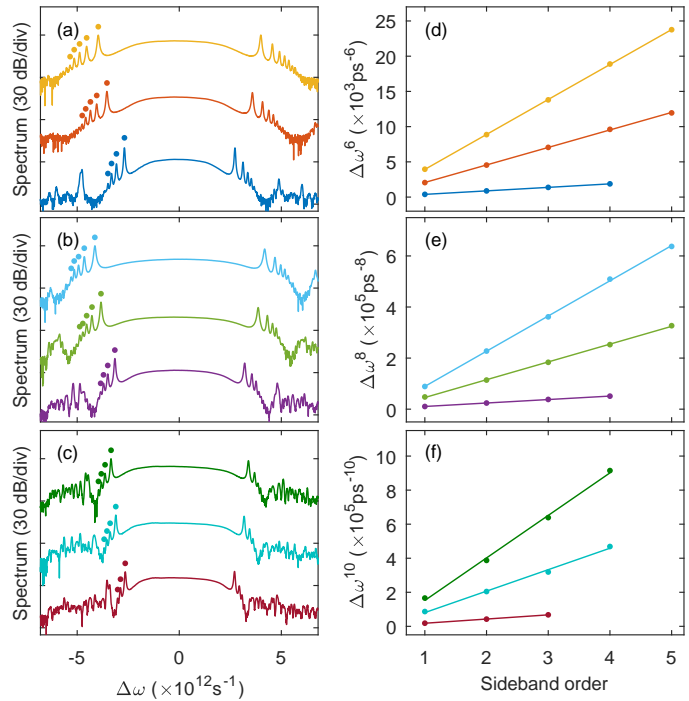}
\caption{Measured PHEOD soliton output spectra for different dispersion orders. (a) Sextic PHEOD soliton spectra for $\beta_6 =-50$ (yellow), $\beta_6=-100$ (orange), and $\beta_6=-500~\rm{ps^6/km}$ (blue). (b) Octic PHEOD soliton spectra for $\beta_8 = -100$ (cyan), $\beta_8=-200$ (green) and $\beta_8=-1000~\rm{ps^8/km}$ (purple). (c) Decic PHEOD soliton spectra for $\beta_{10} = -5000$ (green), $\beta_{10}= -10\times10^3$ (cyan), and $\beta_{10}=-50\times10^3~{\rm ps^{10}/km}$ (red). Coloured circles show the $k^{\rm th}$ power of the measured sidebands positions versus sideband order for the (d) sextic; (e) octic; and (f) decic PHEOD soliton spectra. The solid lines correspond to linear fits.}
\label{sidebands_fig}
\vskip-1mm
\end{figure}

\subsection{Sideband analysis}

To confirm the nature of the cavity's linear dispersion, we analyzed the position of the spectral sidebands of the emitted pulses. These dispersive waves arise from the constructive interference between the solitons and the linear waves emitted by the soliton while it propagates inside the cavity \cite{Kelly_1992, Runge_2020}. Constructive interference occurs when $\beta_{\rm sol} - \beta_{\rm lin} = 2m\pi/L$ where $m$ is a positive integer. For $k^{th}$ order dispersion the linear waves satisfy $\beta_{\rm lin} = -|\beta_k|(\omega-\omega_0)^k/k!$, while the PHEOD solitons have a constant dispersion across its entire bandwidth of $\beta_{\rm sol} = C_k|\beta_k|/\tau^k$ \cite{Akhmediev_1995, Tam_2019}, where $C_k$ are constants of order unity that depend on the dispersion order. Thus, we find that the spectral position of the $m^{th}$ spectral sideband $\omega_m$ is given by \begin{equation}
   \omega_m = \pm{1\over\tau}\left[k! \left({m\, \pi\,\tau^k\over |\beta_k|L}-C_k\right)\right]^{1/k}.
\label{sideband_eq}
\end{equation}
Following this argument, it is straightforward to show that for a pure $k^{th}$ order dispersion soliton, the $k^{th}$ power of two consecutive sidebands is constant and given by $2\pi k!/(|\beta_k|L)$, irrespective of the value of $C_k$. To check this prediction, for each dispersion order we measured the output spectrum for three different values of the dispersion coefficient $\beta_k$ and we measured the spectral positions of the low frequency sidebands.

The results of these measurements are shown in Fig.~\ref{sidebands_fig}. In Fig.~\ref{sidebands_fig}(a)-(c), we show three measured spectra for each dispersion order $k$. The sextic PHEOD soliton spectra for three different values of $\beta_6$ are shown in Fig.~\ref{sidebands_fig}(a). Corresponding results for octic and decic PHEOD solitons are shown in Fig.~\ref{sidebands_fig}(b) and (c), respectively. The circles mark the spectral positions of the low frequency sidebands. The $k^{th}$ power of these measured positions as a function of the sideband order for the nine PHEOD soliton spectra are shown in Fig.~\ref{sidebands_fig}(d)-(f). In all cases the spacings follow a linear relationship as expected. The predicted and measured spectral spacing for all the spectra shown in Fig.~\ref{sidebands_fig}(a)-(c) are summarized in Table~\ref{sideband_table}. Note that the experimental values agree within $4\%$ to the corresponding expected values calculated from \eqref{sideband_eq} and based on the net-cavity dispersion that was applied by the pulse-shaper. Since taking a high power of a dataset amplifies the noise, the agreement between the measured and expected results is remarkable, confirming the type and magnitude of the cavity dispersion. 

\begin{table}[htbp]
\begin{tabular}{|c|c|c|c|}
\hline
Dispersion & Applied $\beta_k$ & Predicted & Measured \\
order $k$ & (${\rm ps}^k{\rm/km}$) & spacing (${\rm ps}^{-k}$) & spacing (${\rm ps}^{-k}$)\\ \hline
  & -50 & 4.98$\times10^3$ & 4.96$\times10^3$ \\
6 & -100 & 2.49$\times10^3$ &  2.48$\times10^3$ \\ 
  & -500 & 4.98$\times10^2$ & 4.97$\times10^2$ \\ \hline
& -100 & 1.39$\times10^5$ & 1.38$\times10^5$ \\
8 & -200 & 6.97$\times10^4$ &  6.96$\times10^4$ \\ 
& -1000 & 1.39$\times10^4$ &  1.36$\times10^4$ \\ \hline
& -5000 & 2.51$\times10^5$ & 2.49$\times10^5$ \\
10 & -10$\times10^3$ & 1.25$\times10^5$ &  1.26$\times10^5$ \\ 
& -50$\times10^3$ & 2.51$\times10^4$ &  2.46$\times10^4$ \\ \hline
\end{tabular}
\caption{Predicted (from \eqref{sideband_eq}) and measured sideband spacing values for different values of applied dispersion $k$.}
\label{sideband_table}
\end{table}

\subsection{Energy-width scaling}

\begin{figure}[hbt]
\centering
\includegraphics[clip = 90mm]{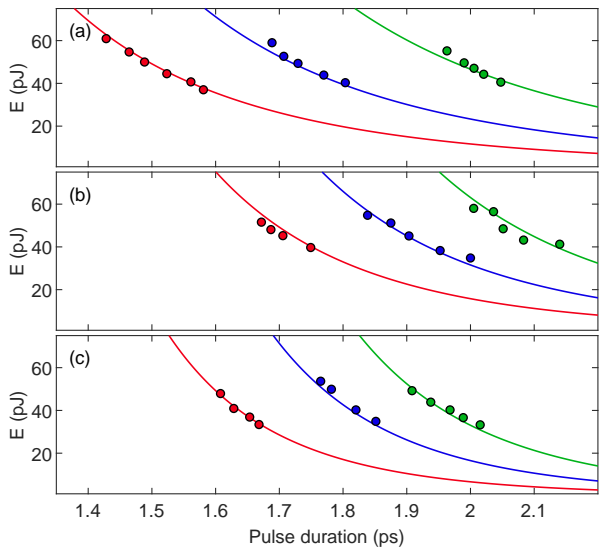}
\vskip 0mm
\caption{Measurement energy-width scaling properties of the pure high-order dispersion Kerr solitons. The circles mark the measured pulse energy $E$ versus pulse duration. (a) Pure-sextic soliton energy for $\beta_6 = -500$ (red), $\beta_6 = -1000$ (blue) and $\beta_6 = -2000~\rm{ps^6/km}$ (green). (b) Pure-octic soliton energy for $\beta_8 = -15\times10^3$ (red), $\beta_8 = -30\times10^3$ (blue) and $\beta_8 = -60\times10^3~\rm{ps^8/km}$ (green). (c) Pure-decic soliton energy for $\beta_{10} = -2\times10^5$ (red), $\beta_{10} = -5\times10^5$ (blue) and $\beta_{10} = -1\times10^6~\rm{ps^{10}/km}$ (green). The solid curves are fits for (a) pure-sextic; (b) -octic; and (c) -decic solitons from \eqref{energy_eq}.}
\label{energy_fig}
\vskip-1mm
\end{figure}

Finally, we study the energy-width scaling relationship of the PHEOD solitons. Following the scaling argument of \eqref{scaling_eq} and by dimensional analysis, we find the energy-width scaling relation of pure high-order dispersion Kerr solitons for $k^{th}$ order of dispersion 
\begin{equation}
   E_k = \frac{M_k|\beta_k|}{\gamma\tau^{k-1}},
\label{energy_eq}
\end{equation}
where $M_k$ is a constant found numerically. For the sextic, octic and decic PHEOD solitons, we found that $M_6 = 0.94$, $M_8 = 0.16$ and $M_{10} = 0.018$, respectively. To confirm this prediction, we measured the output pulse energy as a function of the pulse duration for three different values of dispersion $\beta_k$, for each order of dispersion considered. Concretely, we adjusted the pump power in the laser cavity and measured the output pulse energy after deducting the portion of energy in the spectral sidebands by integrating the optical spectrum. The results of these measurements for $k = 6$, $8$ and $10$ are shown in Fig.~\ref{energy_fig}. The circles show the measured pulse energies versus the pulse duration ($\tau$) for three different values of dispersion $\beta_k$ for each dispersion order $k$. All results are in good agreement with \eqref{energy_eq} once we account for the output coupling and the insertion loss of the pulse-shaper. This shows that the pulse energy $E \propto~\tau^{-(k-1)}$, consistent with \eqref{energy_eq} and confirm that these pure high-order dispersion solitons follow a different energy-width scaling relation that could be used for the generation of ultrashort optical pulses with high energy.

\section{Conclusion and discussion}

We report the experimental discovery of an entire family of optical solitons arising from the balance between SPM and higher-order dispersion. One can consider conventional optical solitons to be the lowest-order member of this family of PHEOD solitons. All these pulses fundamentally arise from similar physical effects: the nonlinearity generates frequencies on the pulses' leading and trailing edges, and these shift towards the pulse center under the effect of dispersion. Our investigation combines numerical results following from solving \eqref{eq:A}, and experimental results obtained using a fiber laser. The fiber laser incorporates a spectral pulse-shaper which is used to apply a large negative pure high, even-order dispersion \cite{Runge_2020}. We find that the numerical and experimental results are in very good agreement. 

As the order of dispersion increases, the soliton spectra become increasingly flat. In addition to this, the peak power of the pulses increases as $\tau^{-k}$, where $k$ is the dispersion order, so that the energy increases as  $\tau^{-(k-1)}$. These features could find applications in high-energy laser systems. Alternatively, the spectral flatness of PHEOD solitons could be used to generate frequency combs with small tooth power variations \cite{TaheriMatsko_2019, Huang_2015}.

Apart from possible laser applications, we have shown that our setup provides a powerful tool to open up new routes for the generation and study of a wide range of novel optical pulses \cite{Lourdesamy_2020}. This approach allows for the simple yet precise tailoring of the net-cavity dispersion and the enhancement of any high-order dispersion effects, which is currently impossible through conventional waveguide dispersion engineering \cite{Blanco_Redondo_2016}. While we only demonstrate PHEOD solitons up to the $10^{th}$ order of dispersion, we emphasize that our approach is only limited by the specifications of the pulse-shaper. This limitation could be overcome by using a device with higher spectral resolution and bandwidth so enabling the generation of PHEOD solitons of order higher than $10^{th}$.  

In addition to direct quantitative insights on novel soliton pulses provided by our experiments, the approach itself is expected to become an established tool for the generation and study of ultrafast pulses \cite{Runge_2020, Lourdesamy_2020}. We expect our results to stimulate future investigations and discoveries in other areas of physics, engineering and applied mathematics.


\section*{Funding Information} 
Australian Research Project (ARC) Discovery Project (DP180102234); 
Asian Office of Aerospace R\&D (AOARD) grant (FA2386-19-1-4067). 

\section*{Disclosures}
A.F.J.R., D.D.H., C.M.de S. and A.B.-R. have submitted a provisional patent application based on the ideas presented in this work.

\bibliography{PQS_bib}
\end{document}